\documentclass{article}

\begin{document}
 
\author{Q. Ho-Kim$^1$ and X. Y. Pham$^2$ \and $^1$\textit{Department of Physics,
Universit\'{e} Laval} \and \textit{Ste-Foy (QC) Canada G1K 7P4} \and {%
e-mail: qhokim@phy.ulaval.ca} \and $^2$\textit{Laboratoire de Physique
Th\'{e}orique et Hautes Energies} \and \textit{Universit\'{e}s Pierre et
Marie Curie \& Denis Diderot} \and \textit{F-75252 Paris Cedex 05 France}
\and {e-mail: pham@lpthe.jussieu.fr}}
\title{One-loop flavor changing electromagnetic transitions}
\date{}
\maketitle

\begin{abstract}
We discuss the effect of the external fermion masses in the flavor-changing
radiative transitions of a heavy fermion (quark or lepton) to a lighter
fermion at the one-loop level, and point out an often overlooked crucial
difference in the sign of a charge factor between transitions of the down
type $s\rightarrow d\gamma $ and the up type $c\rightarrow u\gamma $. We
give formulas for the $F\rightarrow f\gamma $ effective vertex in various
approximations and the exact formula for $t\rightarrow c\gamma $ and $\tau
\rightarrow \mu \gamma $.


\end{abstract}

\newpage
\section{INTRODUCTION}

Flavor-changing radiative transitions $F\rightarrow f\gamma $ are processes
in which a fermion (quark or lepton) undergoes a flavor change accompanied
by the emission of a real or virtual photon. They are the results of an
interplay of the weak and electromagnetic interactions at the loop level,
often enhanced by QCD effects. As such, they have considerable theoretical
and experimental interest, not only because they provide excellent tests of
the standard model but also because they hold the promise of being sensitive
to new physics. They are operating in many different situations, for example
in radiative weak decays of hyperons, such as $\Sigma ^{+}\rightarrow
p\gamma $ and $\Xi ^0\rightarrow \Lambda \gamma $, in rare $B$ meson decays (%
$B\to X_s\gamma $), and in rare $D$ meson decays ($D\to V\gamma $). They may
also occur in rare processes involving leptons, such as $K\rightarrow \pi
e^{+}e^{-}$, $\tau \rightarrow \mu \gamma $, and $\nu _1\rightarrow \nu
_2e^{+}e^{-}$.

At the quark or lepton level, all flavor-changing electromagnetic
transitions may be divided into two categories: one involving the upper
components of weak isospin doublets and the other their lower components, as
shown in Table \ref{LeQu}. In one loop processes, the initial fermion $F$
creates an intermediate state of a boson and a fermion $f_\ell $ which,
after emitting a photon, returns to a fermionic state $f$ as depicted in
Figs. 1 and 2 for an example of quark transition.

\medskip 
{\begin{table}[tbp] \centering} 
\begin{tabular}{lll}
\hline\hline
$F$ & $f$ & $f_\ell $ \\ \hline
$\nu _1$ & $\nu _2$ & $e^{-},\mu^-,\tau^-$ \\ 
$c$ & $u$ & $d,s,b$ \\ 
$t$ & $c(u)$ & $d,s,b$ \\ \hline
$\tau^- $ & $\mu^- (e^-)$ & $\nu _j, j=1,2,3$ \\ 
$s$ & $d$ & $u,c,t$ \\ 
$b$ & $s(d)$ & $u,c,t$ \\ \hline\hline
\end{tabular}
\caption{Elementary flavor-changing radiative processes.\label{LeQu}}%
{\end{table}}

The amplitudes of these penguin processes have been evaluated many years
ago \cite{{inami},{desh}}. In particular, a simple analytic
formula obtained by Inami and Lim \cite{inami} has often served as the
starting point for several more advanced calculations of QCD corrections 
\cite{berto}--\cite{burd} or properties of detailed particle models
\cite{burd}--\cite{fajf}. The interested reader is referred
to a recent review \cite{buras} for further discussion. 

Unfortunately several
authors have overlooked the fact that this formula applies only to
amplitudes of the down-type $sd\gamma $, which were the primary object of
interest to Inami and Lim \cite{inami}, and have used it unchanged to study
processes of the up-type $cu\gamma $. As we shall see, this oversight will
cause an error in sign and magnitude (by a factor of five) in the amplitudes
of the up-type transitions. Moreover, in the down-type $b\rightarrow s\gamma $
transition, the intermediate state is dominated by the contribution of the
top quark and therefore it is well justified to neglect the effects of the
external fermion masses as it was done before \cite{inami}, but it is
doubtful that such an approximation will hold in other cases, especially in $%
c\rightarrow u\gamma $ and $\tau \rightarrow \mu \gamma $, not to mention $%
t\rightarrow c\gamma $, in which the mass of the initial particle is
comparable or larger than any internal (boson or quark) mass.

In this paper, we will correct the above-mentioned shortcomings and proceed
on to examine the effect of the external fermionic masses in flavor-changing
radiative transitions. After writing down the rules and conventions of
calculations in Sec. II, we will evaluate in Sec. III the effective vertex
for $F\rightarrow f\gamma $ assuming first small but non-vanishing external
fermionic masses. Although the methods we follow are applicable to both
quarks and leptons, we refer specifically to quarks for convenience. We
calculate the inclusive rates for $q_i\rightarrow q_j\gamma $ and examine
the effects of the external masses on the decay rates in some detail; we
also give an estimate of the QCD-corrected branching ratio of $c\rightarrow
u\gamma $. In Sec. IV, we derive an exact formula, with $m_c=0$ but
non-vanishing $m_t$, for the effective $t\rightarrow c\gamma $ vertex and
demonstrate the importance of the initial quark mass. We conclude with
Sec.~V.

\section{FEYNMAN RULES \\ IN THE 't HOOFT--FEYNMAN GAUGE}

The standard electroweak model formulated in any renormalizable gauge $R_\xi 
$ contains, besides the usual physical particles (the photon $A_\mu $,
bosons W$_\mu ^{\pm }$, and quarks q$_i$), the unphysical would-be Goldstone
scalar bosons $\varphi ^{\pm },\,\varphi ^0$ coming from the spontaneous
symmetry breaking (as well as the Faddeev--Popov ghosts $\chi^{\pm},\, \chi^0$ which, however, do
not contribute to processes considered here because the external particles
are fermions). In writing down the following Feynman rules we use the
convention that defines \emph{ingoing} momenta for particles entering the
vertex and \emph{outgoing} momenta for particles leaving the vertex. Quark
flavors will be denoted by $i$ or $j$, and the left and right projection
operators by $L=(1-\gamma _5)/2$ and $R=(1+\gamma _5)/2$ .

The rules for the elementary vertices are 
\begin{eqnarray*}
q_i &\rightarrow &q_i\,A_\mu \,:-ieQ_i\gamma _\mu \quad \,\mathrm{with}\,\;
Q_i=\frac 23\;(u),\,-\frac 13\,\;(d)\,, \\
W_\lambda ^{\pm }(p) &\rightarrow &W_\nu ^{\pm }(p^{\prime })\,A_\mu
\,:+i\,eQ_W[(p+p^{\prime })_\mu \,g_{\lambda \nu }+(p^{\prime }-2p)_\nu
\,g_{\lambda \mu } \\
&&\hspace*{2cm}+(p-2p^{\prime })_\lambda \,g_{\mu \nu }\,]\quad \mathrm{with}%
\;Q_W=\pm 1(W^{\pm })\,, \\
\varphi ^{\pm }(p) &\rightarrow &\varphi ^{\pm }(p^{\prime })\,A_\mu \,:-%
\mathrm{\,}ieQ_S\,(p+p^{\prime })_\mu \quad \,\mathrm{with}\,\;Q_S=\pm
1\,(\varphi ^{\pm })\,, \\
W_\nu ^{\pm } &\rightarrow &\varphi ^{\pm }\,A_\mu :+\,i\,e\,M_W\,g_{\mu \nu
}\,, \\
q_i &\rightarrow &q_jW_\mu ^{\pm }:\,\frac{-ig}{\sqrt{2}}\gamma _\mu
\,LU_{ji}^{(I_i)}\,, \\
q_i &\rightarrow &q_j\varphi ^{\pm }:\,\frac{-ig}{\sqrt{2}\,M_W}%
\,I_i\,(m_jL-m_iR)\,U_{ji}^{(I_i)}\,.
\end{eqnarray*}

The symbol $I_i$ which appears in the last two rules is related to the weak
isospin of the initial quark. For $I_i=+1$ ($u$-type quark) the CKM matrix
element is $U_{ji}^{(+)}=V_{ij}^{*}$, and for $I_i=-1$ ($d$-type quark) $%
U_{ji}^{(-)}=V_{ji}$. Note that $I_i=Q_W$ (or $Q_S$) and thus, with the
exception of the $\varphi AW$ and $qqW$ vertices, all other vertices depend
on the sign of the charge of the quark or boson involved. Finally, $e$ is
the positive unit of the electric charge, $e>0$.

The relevant boson propagators in the 't Hooft--Feynman ($\xi =1$) gauge are 
\begin{eqnarray*}
\frac i{p^2-M_W^2}\quad &&\mathrm{(scalar)}, \\
\frac{-i\,g_{\mu \nu }}{p^2-M_W^2}\quad &&(W\mathrm{\ boson}).
\end{eqnarray*}

\section{CONTRIBUTIONS TO \\ THE EFFECTIVE VERTEX}

In the lowest order, i.e. at the one-loop level, the processes in which a
quark undergoes a flavor change accompanied by the emission of a real or
virtual photon are represented by the diagrams in Figs.~1(a)--(d) and
Figs.~2(a)--(b). In the first group, the photon is emitted by a scalar or a
vector boson in the intermediate state, whereas in the second it arises from
an internal fermion. As we are eventually interested in real, transversally
polarized photon emission, the transition should take place in a magnetic
mode characterized by the operator $i\sigma _{\mu \nu }q^\nu $, and we need
not concern ourselves with diagrams in Fig. 3 since they are all
proportional to $\gamma _\mu $.

We perform our calculations in $n$ dimensions and let $n=4-2\omega $. For
convenience we also introduce an arbitrary scale parameter with the
dimension of mass $\mu $ so that the coupling constant $g$ remains
dimensionless even in arbitrary $n$ dimensions. The kinematic variables to
be used are defined in Fig.~4. In particular, the initial quark has momentum 
$P$ and mass $M$; the final quark has momentum $p$ and mass $m$. The
internal quark, with mass $m_\ell $, may have flavors $u,\, c$, or $t$ for an
initial quark of the $d$-type, and $s, \,d$, or $b$ for an initial quark of the
$u$-type. It is understood that the various partial or total vertex operators
obtained in the following for $q_i\rightarrow \,q_f\,\gamma $ are to be
inserted between the initial state $u_i(P)$ and the final state $\bar{u}%
_f(p)\varepsilon ^{\mu \,*}(q)$, where $u_i$ and $u_f$ are the fermion
spinors, and $\varepsilon ^\mu (q)$ is the photon polarization vector with
momentum $q=P-p$. A summation over all allowed quark flavors $\ell $ and an
integration over the loop $n$-dimensional $k$ momentum are both implicit.

From the rules given above, the expression corresponding to Fig.~1(a) is 
\begin{equation}
iT_{1\mathrm{a}}^\mu =\mu ^{2\omega }\left( \frac{-ig}{\sqrt{2}}\right)
^2\lambda _\ell \,\gamma ^\rho L\frac i{\not{k}-m_\ell }\,\gamma ^\sigma L%
\frac{(-i)^2(ie\,Q_W\,X^\mu {}_{\rho \sigma })}{%
[(p-k)^2-M_W^2][(P-k)^2-M_W^2]},  \label{ampl1a}
\end{equation}
where the following shorthand notations have been used: 
\begin{equation}
X_{\mu \rho \sigma }=(P+p-2k)_\mu \,g_{\rho \sigma }+(P+k-2p)_\sigma
\,g_{\rho \mu }+(p+k-2P)_\rho \,g_{\mu \sigma }
\end{equation}
and 
\begin{eqnarray}
i &=&u\ \mathrm{type:}\qquad Q_W=+1,\;\lambda _\ell =V_{f\ell}V_{i\ell }^{*} 
\nonumber \\
i &=&d\ \mathrm{type:}\qquad Q_W=-1,\;\lambda _\ell =V_{\ell f}^{*}V_{\ell i}
\;.\label{baa}
\end{eqnarray}
It is convenient to put (\ref{ampl1a}) in the form 
\begin{equation}
iT_{1\mathrm{a}}^\mu =\frac{-g^2e}2\mu ^{2\omega }Q_W\lambda _\ell \frac{N_{1%
\mathrm{a}}^\mu }{D_1}
\end{equation}
with the following expressions for the numerator and denominator: 
\begin{eqnarray}
N_{1\mathrm{a}}^\mu &=&\gamma ^\rho L\,(\not{k}+m_\ell )\,\gamma ^\sigma
L\,X^\mu {}_{\rho \sigma }\,,  \nonumber \\
D_1 &=&(k^2-m_\ell ^2)\left[ (p-k)^2-M_W^2\right] \left[
(P-k)^2-M_W^2\right] \,.
\end{eqnarray}

Corresponding to Fig.~1(b) is the transition operator 
\begin{eqnarray}
iT_{1\mathrm{b}}^\mu &=&\mu ^{2\omega }\lambda _\ell \left[ \frac{igQ_S}{%
\sqrt{2}M_W}(mL-m_\ell R)\right] \,\frac i{\not{k}-m_\ell }\,\left[ \frac{%
-igQ_S}{\sqrt{2}M_W}(m_\ell L-MR)\right] \,  \nonumber \\
&&\times \frac{i^2(-ieQ_S)(P+p-2k)^\mu }{[(p-k)^2-M_W^2][(P-k)^2-M_W^2]} 
\nonumber \\
&=&\frac{-g^2e}2\mu ^{2\omega }\,Q_S\lambda _\ell \frac{N_{1\mathrm{b}}^\mu 
}{D_1}\,.\label{sau}
\end{eqnarray}
From now on, if the initial quark flavor $q_i$ is of the $u$-type, $Q_S=+1$; otherwise $%
Q_S=-1$. The $\lambda_\ell$ always follows the rules of Eq. (\ref{baa}). The expression for the denominator $D_1$ remains the same as
defined above, whereas in the numerator one has 
\begin{equation}
N_{1\mathrm{b}}^\mu =\frac 1{M_W^2}(mL-m_\ell R)(\not{k}+m_\ell )(m_\ell
L-MR)(P+p-2k)^\mu \,.
\end{equation}

For Fig.~1(c), the Feynman rules yield 
\begin{eqnarray}
iT_{1\mathrm{c}}^\mu &=&\mu ^{2\omega }\lambda _\ell \left( \frac{-ig}{\sqrt{%
2}}\gamma _\mu L\right) \frac i{\not{k}-m_\ell }\,\left[ \frac{-igQ_S}{\sqrt{%
2}M_W}(m_\ell L-MR)\right]  \nonumber \\
&&\times \frac{(ieM_W)(-i)(i)}{[(p-k)^2-M_W^2][(P-k)^2-M_W^2]}  \nonumber \\
&=&\frac{-g^2e}2\mu ^{2\omega }\,Q_S\lambda _\ell \frac{N_{1\mathrm{c}}^\mu 
}{D_1}\,,
\end{eqnarray}
with the numerator given by 
\begin{equation}
N_{1\mathrm{c}}^\mu =\gamma ^\mu L(\not{k}+m_\ell )(MR-m_\ell L)\,.
\end{equation}

The diagram in Fig.~1(d) is similar to that in Fig.~1(c), with the scalar
and vector bosons exchanging roles in the intermediate state. The
corresponding vertex operator is 
\begin{eqnarray}
iT_{1\mathrm{d}}^\mu &=&\mu ^{2\omega }\lambda _\ell \,\left[ \frac{+igQ_{S}%
}{\sqrt{2}M_{W}}(mL-m_\ell R)\right] \frac{i}{\not{k}-m_\ell }\left( \frac{%
-ig}{\sqrt{2}} \gamma_\mu L\right)  \nonumber \\
&&\times \frac{(ieM_{W})(i)(-i)}{[(p-k)^2-M_{W}^2][(P-k)^2-M_{W}^2]} 
\nonumber \\
&=&\frac{-g^2e}2\mu ^{2\omega }\,Q_{S}\lambda _\ell \frac{N_{1\mathrm{d}%
}^\mu }{D_1}\,,
\end{eqnarray}
with the following expression in the numerator 
\begin{equation}
N_{1\mathrm{d}}^\mu =(mL-m_\ell R)(\not{k}+m_\ell )\gamma ^\mu L \,.
\end{equation}

The photon may be emitted also from the intermediate quark as shown in
Figs.~2(a)--(b). The diagram in Fig.~2(a) represents the transition operator 
\begin{eqnarray}
iT_{2\mathrm{a}}^\mu &=&\mu ^{2\omega }\lambda _\ell \left( \frac{-ig}{\sqrt{%
2}}\gamma ^\rho L\right) \frac{i}{\not{p}- \not{k}-m_\ell }\left( -ieQ_\ell
\gamma ^\mu \right)  \nonumber \\
&&\times \frac{{i}}{\not{\!P}-\not{k}-m_\ell }\left( \frac{-{i}g}{\sqrt{2}}%
\gamma _\rho L\right) \frac{-i}{k^2-M_{W}^2}  \nonumber \\
&=&\frac{-g^2e}2\mu ^{2\omega }\,Q_\ell \lambda _\ell \,\frac{N_{2\mathrm{a}%
}^\mu }{D_2}\,,
\end{eqnarray}
where $N_{2\mathrm{a}}^\mu $ and $D_2$ stand for 
\begin{eqnarray}
N_{2\mathrm{a}}^\mu &=&\gamma ^\rho (\not{p}-\not{k}+m_\ell )\gamma ^\mu (%
\not{\!P}-\not{k}+m_\ell )\gamma _\rho L \,,  \nonumber \\
D_2 &=&(k^2-M_{W}^2)\left[ (p-k)^2-m_\ell ^2\right] \left[ (P-k)^2-m_\ell
^2\right]\,.
\end{eqnarray}
Similarly for Fig. 2(b), we have 
\begin{eqnarray}
iT_{2\mathrm{b}}^\mu &=&\mu ^{2\omega }\lambda _\ell \left[ \frac{+igQ_{{S}}%
}{\sqrt{2}M_{{W}}}(mL-m_\ell R)\right] \frac{i}{\not{p}-\not{k}-m_\ell }%
\left( -ieQ_\ell \gamma ^\mu \right)  \nonumber \\
&&\times \frac{{i}}{\not{\!P}-\not{k}-m_\ell }\,\left[ \frac{-igQ_{{S}}}{%
\sqrt{2}M_{{W}}}(m_\ell L-MR)\right] \frac{i}{k^2-M_{W}^2}  \nonumber \\
&=&\frac{-g^2e}2\mu ^{2\omega }\,Q_\ell \lambda _\ell \,\frac{N_{2\mathrm{b}%
}^\mu }{D_2}\,,
\end{eqnarray}
together with the expression for the numerator 
\begin{equation}
N_{2\mathrm{b}}^\mu =\frac 1{M_{W}^2}(mL-m_\ell R)(\not{p}-\not{k}+m_\ell
)\gamma ^\mu (\not{\!P}-\not{k}+m_\ell )(m_\ell L-MR)\,.
\end{equation}

Thus, all partial contributions to the effective radiative vertex may be
written in the form 
\begin{equation}
iT_{p\alpha }^\mu =\frac{-g^2e}2\mu ^{2\omega }\,Q_p\lambda _\ell \,\frac{%
N_{p\alpha }^\mu }{D_p}\qquad (p=1,2;\,\alpha ={a,b,}\ldots )
\label{partial}
\end{equation}
with the charge factor 
\begin{equation}
Q_1=Q_B\,(Q_{W}\,\mathrm{or\,}Q_{S}),\qquad Q_2=Q_\ell .  \label{charge}
\end{equation}

After Feynman parameterization of the denominators, the integration over $n$%
-dimensional momentum space is performed in the standard way. Consider for
example $T_{1\mathrm{a}}^\mu $. The terms independent of the integrated
momentum $k$ are evaluated as follows: 
\begin{eqnarray*}
\mu ^{2\omega }\int \frac{\mathrm{d}^nk}{(2\pi )^n}\frac 1{D_1} &=&\mu
^{2\omega }\int \frac{\mathrm{d}^nk}{(2\pi )^n}\,\int_0^1\!\mathrm{d}%
x\int_0^{1-x}\!\mathrm{d}y\frac 2{[a(1-x-y)+bx+cy]^3} \\
&=&\frac{-i}{(4\pi )^2}\int_0^1\!\mathrm{d}x\int_0^{1-x}\!\mathrm{d}y\frac
1{M_{{W}}^2\,\Lambda _1}\left( \frac{4\pi \,\mu ^2}{-M_{{W}}^2\,\Lambda _1}%
\right) ^\omega \Gamma (1+\omega )
\end{eqnarray*}
for $a=(p-k)^2-M_{{W}}^2$, $b=k^2-m_\ell ^2$, $c=(P-k)^2-M_{{W}}^2$, and 
\begin{eqnarray*}
M_{{W}}^2\,\Lambda _1 &=&M_{{W}}^2\,(1-x)+m_\ell ^2x-P^2xy-p^2xz-q^2yz\,, \\
z &=&1-x-y\,.
\end{eqnarray*}
Terms with explicit $k$-dependent integrands are handled in a similar way.
With the summation over the internal quarks now explicit, the $k$ 
integration therefore carries Eq.~(\ref{ampl1a}) into 
\begin{eqnarray}
iT_{1\mathrm{a}}^\mu &=&\frac{-g^2e}2\,Q_{{W}}\,\mu ^{2\omega }\sum_\ell
\lambda _\ell \int \frac{\mathrm{d}^nk}{(2\pi )^n}\frac{N_{1\mathrm{a}}^\mu 
}{D_1}  \nonumber \\
&=&\frac{ig^2e}{32\pi ^2M_{{W}}^2}\,Q_{{W}}\sum_\ell \lambda _\ell \int_0^1\!%
\mathrm{d}x\int_0^{1-x}\!\mathrm{d}y\frac 1{\,\Lambda _1}  \nonumber \\
&&\times R\,\left\{ [-4xP^\mu +2(1-2z)q^\mu ][(1-x)\not{\!P}-z\not{q}]\right.
\nonumber \\
&&+(1-x)(\not{q}\gamma ^\mu \not{\! P}-\not{\! P}\gamma ^\mu \not{q})+(-x\not%
{\!P}-z\not{q})[(1-x)\not{\! P}-z\not{q}]\gamma ^\mu  \nonumber \\
&&+\gamma ^\mu [(1-x)\not{\! P}-z\not{q}][-x\not{\! P}+(1-z)\not{q}]+\left.
4(1-n)\gamma ^\mu \mathcal{F}_1\right\} \,.
\end{eqnarray}
Here $\mathcal{F}_1$ comes from the singular part of the quadratic $k$%
-dependent integrand. This result will considerably simplify when the
external particles go on shell, which is all we need. When the initial and
final quarks are on the mass shell, we may use the Dirac equations ${\not{\!
P}}u_i(P)=Mu_i(P)$ and $\bar{u}_f(p)\!{\not{p}}=m\bar{u}_f(p)$ to express
the operator $T_{1\mathrm{a}\,\mu }$ in terms of the four-vectors $P_\mu $, $%
q_\mu $, and $\gamma _\mu $ or, alternatively, $i\sigma _{\mu \nu }q^\nu $, $%
q_\mu $, and $\gamma _\mu $, the two bases being related by 
\begin{eqnarray}
2P_\mu R &=&R(i\sigma _{\mu \nu }q^\nu +q_\mu +M\gamma _\mu )+Lm\gamma _\mu
\,,  \nonumber \\
2P_\mu L &=&L(i\sigma _{\mu \nu }q^\nu +q_\mu +M\gamma _\mu )+Rm\gamma _\mu
\,.
\end{eqnarray}
These relations are understood as being sandwiched between $\bar{u}(p)$ and $%
u(P)$. The induced complete $q_i q_f\gamma $ vertex will then assume the
general form 
\begin{equation}
T_\mu \sim R(q^2\gamma _\mu -q_\mu \not{q})F+i\sigma _{\mu \nu }q^\nu
(RMG+LmH)\,,
\end{equation}
where $F$, $G$, and $H$ are appropriate form factors. For an on-shell
photon, $q^2=0$, we may set $q_\mu \varepsilon ^\mu (q)=0$, and the
coefficient of $F$ should vanish. The singular terms, such
as $\mathcal{F}_1$, from all diagrams likewise completely cancel out \cite
{hoki}. Hence, with real photon radiative processes in mind, we may drop all
such terms in the individual contributions. When the above simplifications
are applied, $T_{1\mathrm{a}}^\mu $ reduces on the mass shell to 
\begin{eqnarray}
T_{1\mathrm{a}}^\mu &=&\frac{e\,G_{\mathrm{F}}}{4\sqrt{2}\pi ^2}Q_{{\mathrm{W%
}}}\sum_\ell \lambda _\ell \int_0^1\mathrm{d}x\int_0^{1-x}\!\mathrm{d}y\frac
1{\,\Lambda _1}\,  \nonumber \\
&&\times \left\{ RM[1-x+y(1-2x)]+Lm[1-x+z(1-2x)]\right\} \,i\sigma ^{\mu \nu
}q_\nu \,,
\end{eqnarray}
where $G_{\mathrm{F}}/\sqrt{2}=g^2/(8M_{W}^2)$ and 
\begin{equation}
\,\Lambda _1=\,1-x+r_\ell \,x-r_i\,xy-r_f\,xz \label{haihai}
\end{equation}
with $r_\ell =m_\ell ^2/M_{W}^2$, $r_i=M^2/M_{W}^2$, and $r_f=m^2/M_{W}^2$.
The contributions from all the other diagrams may be reduced to this
standard form.

Let us sum up the partial contributions in each group ($p=1,2$) of diagrams
to get the final transition operators 
\begin{equation}
T_p^\mu =\frac{e\,G_{\mathrm{F}}}{4\sqrt{2}\pi ^2}Q_p\sum_\ell \lambda _\ell
\,[RMG_p(r_\ell )+LmH_p(r_\ell )]\,i\sigma ^{\mu \nu }q_\nu \,,
\label{RLAmp}
\end{equation}
where we have used $Q_p$ defined in (\ref{charge}), and introduced 
\begin{eqnarray}
G_p(r_\ell) &=&\int_0^1\!\mathrm{d}x\int_0^{1-x}\!\mathrm{d}y\,\frac 1{\,\Lambda
_p}\,g_p(r_\ell,x,y)\,,  \nonumber \\
H_p(r_\ell) &=&\int_0^1\!\mathrm{d}x\int_0^{1-x}\!\mathrm{d}y\,\frac 1{\,\Lambda
_p}\,h_p(r_\ell,x,y)\qquad (p=1,2)\,,  \label{GHp}
\end{eqnarray}
together with 
\begin{eqnarray}
g_1(r_\ell,x,y) &=& 1-x+z+y(1-2x)+r_\ell x(1-y) -r_fxz\,,  \nonumber \\
h_1(r_\ell,x,y) &=& 1-x+y+z(1-2x)+r_\ell x(1-z)-r_i xy\,,  \nonumber \\
g_2(r_\ell,x,y) &=& -2x(1-y)+r_\ell (-1+x+xy) + r_f xz\,,  \nonumber \\
h_2(r_\ell,x,y) &=& -2x(1-z)+r_\ell (-1+x+xz) + r_i xy\,,
\end{eqnarray}
where $\Lambda_1$ is given by (\ref{haihai}), and $\Lambda_2 =\,x +r_\ell (1-x) -r_i\,xy-r_f\,xz$.   
These expressions display a symmetry between $g_p$ and $h_p$, for a given $p$%
, under simultaneous interchange of $y$, $r_i$ and $z$, $r_f$. They reduce
to 
\begin{eqnarray}
g_1(r_\ell,x,y) &=&2+(r_\ell -2)x-(r_\ell +2)xy -r_fxz\,,  \nonumber \\
h_1(r_\ell,x,y) &=&2-4x+(r_\ell +2)x^2+(r_\ell +2-r_i)xy\,,  \nonumber \\
g_2(r_\ell,x,y) &=&-r_\ell +(r_\ell -2)x+(r_\ell +2)xy+r_f xz\,,  \nonumber \\
h_2(r_\ell,x,y) &=&-r_\ell + 2r_\ell x-(r_\ell +2)x^2-(r_\ell+2-r_i)xy\,.
\end{eqnarray}

An obstacle to a simple analytic expression for the effective vertex is the $%
x$ integration in (\ref{GHp}). However, if both initial and final quark
masses are small compared with the gauge boson masss, i.e. if $r_i,\,r_f\ll
1 $, one may consider making a linear approximation in $r_i$ and $r_f$: 
\begin{eqnarray}
\frac 1{\Lambda _p} &\approx &\frac 1{\Lambda _p^{\!(0)}}+\frac 1{\left(
\Lambda _p^{\!(0)}\right) ^2}\,x(r_i\,y+r_f\,z)\qquad (p=1,2)  \label{denom}
\\
\mathrm{for\;}\Lambda _1^{\!(0)} &=&1-x+r_\ell \,x\,,\quad \Lambda
_2^{\!(0)}=x+r_\ell \,(1-x\,)\,.  \nonumber
\end{eqnarray}
In this approximation of $\Lambda _p,$ the functions $G_p$ and $H_p$ may be
written as 
\begin{eqnarray}
G_p &=&F_p(r_\ell)+r_f\,K_p(r_\ell)+r_i\,L_p(r_\ell)\,,  \nonumber \\
H_p &=&F_p(r_\ell)+r_i\,K_p(r_\ell)+r_f\,L_p(r_\ell) \quad (p=1,2)\,,  \label{Ghp}
\end{eqnarray}
with a symmetry reflecting that between $g_p$ and $h_p$. Here $F_p(r_\ell)$, $K_p(r_\ell)$,
and $L_p(r_\ell)$ are functions of $r_\ell $ only. Writing $g_p=g_p^{(0)}+r_f
g_p^{(1)}$ and $h_p=h_p^{(0)}+r_i h_p^{(1)}$, we see that $F_p$ comes from
the $xy$ integral $\int g_p^{(0)}/\Lambda _p^{\!(0)}$ (or its equal $\int
h_p^{(0)}/\Lambda _p^{\!(0)}\,$), while $K_p$ and $L_p$ arise from various
terms linear in $r_i$ or $r_f$ in $g_p$, $h_p$, or $\Lambda_p$. From (\ref{Ghp}), together the
functions $F_p(r_\ell)$, $K_p(r_\ell)$, and $L_p(r_\ell)$ determine $G_p$ and $H_p$ for $p=1,2$ to
first order in $r_i\,,\,r_f$. They are given by ($d=r-1$) 
\begin{eqnarray}
F_1(r) &=&\displaystyle \frac r{4d^3}(1-5r-2r^2)+\frac{3r^3}{2d^4}\,\ln r\,,
\nonumber \\[0.5ex]
K_1(r) &=&\displaystyle \frac r{72d^5}(13-11r-195r^2+17r^3-4r^4)  \nonumber
\\[0.5ex]
&&+\displaystyle \frac{r^2}{6d^6}(-3+13r+5r^2)\,\ln r,  \nonumber \\[0.5ex]
L_1(r) &=&\displaystyle \frac r{36d^5}(7-65r-141r^2+23r^3-4r^4)  \nonumber \\%
[0.5ex]
&&+\displaystyle \frac{r^3}{3d^6}(13+2r)\,\ln r,  \nonumber \\[0.5ex]
F_2(r) &=&\displaystyle \frac r{4d^3}(-2-5r+r^2)+\frac{3r^2}{2d^4}\,\ln r, 
\nonumber \\[0.5ex]
K_2(r) &=&\displaystyle \frac r{72d^5}(41-237r+39r^2-29r^3+6r^4)  \nonumber
\\[0.5ex]
&&+\displaystyle \frac r{6d^6}(-2+8r+9r^2)\,\ln r,  \nonumber \\[0.5ex]
L_2(r) &=&\displaystyle \frac r{18d^5}(-5-105r+33r^2-16r^3+3r^4)  \nonumber
\\
&&+\displaystyle \frac r{6d^6}(-1+22r+9r^2)\,\ln r\,.  \label{lastKL}
\end{eqnarray}

We have omitted in the above expressions terms that are numerical constants.
Such terms independent of $r_\ell $ will give vanishing contributions as
long as $F_p$, $K_p$, and $L_p$ are used under the sums carried out over all
allowed $\ell $, as in Eq.~(\ref{RLAmp}), for $\sum_\ell \lambda _\ell =0$
by the unitarity of the CKM matrix. 

It is crucial to note that for the down-type transitions such as $s\rightarrow d$ and $b\rightarrow s$, we have $Q_1=-1$, $Q_2=2/3$
and $\ell =1,2,3$ correspond respectively to $u,\, c, \,t$; whereas for up-type transitions $c\rightarrow u$ and $t\rightarrow c$, we have 
$Q_1=+1$, $Q_2=-1/3$, and $\ell = d, \,s,\, b$.

Inami and Lim \cite{inami} have entirely neglected the external masses in their calculations of $G_p$ and $ H_p$
by taking  $r_i= r_f=0$,  so that $G_p = H_p = F_p$ in the general formula Eq.~(\ref{RLAmp}). For the stated
purpose of their work, they have given the explicit expression for $%
-F_1+(2/3)F_2$ valid for the down-type transitions like $s\rightarrow d\gamma $ or $b\rightarrow s\gamma $;
namely, 
\begin{eqnarray}
-F_1+\frac 23F_2 &=&\frac r{12(r-1)^4}\left[ (1-r)(7-5r-8r^2)+6r(2-3r)\ln
r\right] \stackrel{r\to 0}{\rightarrow }\frac 7{12}r\,.  \nonumber \\
&&
\end{eqnarray}
But for the up-type transitions such as $c\rightarrow u\gamma $ (always within the $r_i= r_f =0$ approximation), one 
should use instead the expression 
\begin{eqnarray}
F_1-\frac 13F_2 &=&\frac{-r}{12(r-1)^4}\left[ (1-r)(5-10r-7r^2)+6r(1-3r)\ln
r\right] \stackrel{r\to 0}{\rightarrow }\frac{-5}{12}r\,,  \nonumber \\
&&  \label{right}
\end{eqnarray}
and not the following combination \cite{burd}, \cite{fajf} or its
limiting value \cite{greub} 
\begin{eqnarray}
-F_1-{\frac 13}F_2 &=&{\frac r{12(r-1)^4}}\left[
(1-r)(1-10r-5r^2)-6r(1+3r)\ln r\right] \stackrel{r\to 0}{\longrightarrow }{%
\frac 1{12}}r.  \nonumber \\
&&  \label{wrong}
\end{eqnarray}
As for the leptonic transitions listed in Table 1,
the up-type heavy neutrino decay amplitude $\nu_1 \to \nu_2\gamma\;$  having $Q_1=1,\,Q_2=-1\,$  is associated with the combination \cite{hoki}
\begin{equation}
 F_1-F_2= 
\displaystyle  \frac{3r}{4(r-1)^3} (1-r^2+2r\ln r)\,, 
\label{baba}
\end{equation}
which differs from \cite{pal} by a constant $3/2$. As noted before, when summation over $\ell$ is carried in the amplitude (\ref{RLAmp}), this $r_\ell$-independent constant $3/2$ does not contribute however, due to 
$\sum_\ell \lambda_\ell =0$. 

For the down-type $\tau^{-}\to\mu^{-}\gamma\;(Q_1=-1,\,Q_2=0)$ on the other hand, the corresponding amplitude is simply associated with $\,-F_1 $.

To examine the relative importance of the external masses, we calculate the
decay rates of $q_i\rightarrow q_f\gamma $, using the linear approximations
in $r_i,\,r_f$, as defined in Eqs. (\ref{RLAmp})--(\ref{lastKL}). For this
purpose, it is convenient to define the amplitudes 
\begin{eqnarray}
S_p(r) &=&\frac 12[G_p(r)+\sqrt{r_f/r_i}\,H_p(r)]\,,  \nonumber \\
P_p(r) &=&\frac 12[G_p(r)-\sqrt{r_f/r_i}\,H_p(r)]\,,  \label{Spp}
\end{eqnarray}
so that Eq.~(\ref{RLAmp}), when summed over both groups of diagrams, may be
rewritten as 
\begin{equation}
T_\mu =\frac{e\,G_{\mathrm{F}}\,M_W\sqrt{r_i}}{4\sqrt{2}\,\pi ^2}\left(
A+B\,\gamma _5\right) i\sigma _{\mu \nu }q^\nu \,,  \label{ABAmp}
\end{equation}
where 
\begin{eqnarray}
A &=&\sum_{\ell =1}^3\lambda _\ell \sum_{p=1}^2\,Q_pS_p(r_\ell )=\sum_{\ell
=1}^2\lambda _\ell \sum_{p=1}^2\,Q_p[S_p(r_\ell )-S_p(r_1)]\,, \\
B &=&\sum_{\ell =1}^3\lambda _\ell \sum_{p=1}^2\,Q_pP_p(r_\ell )=\sum_{\ell
=1}^2\lambda _\ell \sum_{p=1}^2\,Q_p[P_p(r_\ell )-P_p(r_1)]\,.
\end{eqnarray}
In the last step we have made use of the unitarity of the CKM matrix to
replace $\lambda _1$ with $-(\lambda _2+\lambda _3)$.  In terms of these amplitudes the decay rate for $%
q_i\rightarrow q_f\gamma $ is given by 
\begin{equation}
\Gamma (q_i\rightarrow q_f\gamma )=\frac{G_{\mathrm{F}}^2\,\alpha }{64\pi ^4}%
M_{W}^5\left( \sqrt{r_i}\right) ^5\left( 1-\frac{r_f}{r_i}\right) ^3\left(
\left| A\right| ^2+\left| B\right| ^2\right) \,.  \label{DecRat}
\end{equation}

The results of our calculations based on Eq.~(\ref{DecRat}) are shown in
Table \ref{Width} together with $\Gamma _0$, the corresponding values of the
width when the factors $r_i$ and $r_f$ in the amplitudes are
neglected. We have used $m_u=5\,$MeV, $m_c=1.5\,$GeV, $m_t=174\,$GeV, $%
m_d=11\,$MeV, $m_s=150\,$MeV, and $m_b=4.9\,$GeV for the quark masses, $%
M_W=80\,$GeV for the $W$ boson mass, and the central values of the CKM matrix
elements as given in the Review of Particle Physics \cite{RevPP}.

We see that, with one exception (the $t\rightarrow
c\gamma $ case), generally $\Gamma \approx \Gamma _0$, i.e.
apart from the kinematic factor $\left( \sqrt{r_i}\right) ^5$, the external
masses can be safely neglected in calculating the width for $q_i\rightarrow
q_f\gamma $. The reason is that the external mass correction functions $K_p$ and $L_p$
enter the amplitudes $S_p$ and $P_p$ respectively accompanied by the mass
factors $\sqrt{r_ir_f}$ and $[(\sqrt{r_i})^3+(\sqrt{r_f})^3]/(\sqrt{r_i}+%
\sqrt{r_f}),$ which are much smaller than $1$, the coefficient of $F_p$
[cf.~Eqs. (\ref{Ghp}) and (\ref{Spp})]. The exceptional case is the $%
t\rightarrow c\gamma $ transition. Since $r_{t}\approx 4.73$, neglecting the
mass of the initial quark is not justified, and even keeping just terms
linear in $r_{t}$ as in Eqs. (\ref{denom}), (\ref{Ghp}) is not enough and
misleading: the approximation has broken down. However, the fact that $%
\Gamma \approx 100\,\Gamma _0$ for this transition clearly indicates the
importance of the effect of the external top mass. In Table \ref{TabMod} we
list the contributions of the intermediate quarks ($d, s$, and $b$) to the
amplitudes $S=S_1-(1/3)S_2$ and $P=P_1-(1/3)P_2$ of the $t\rightarrow
c\gamma $ transition for both $m_{c}=0$ and $m_{c}=1.5\,$GeV. These data
show that the mass of the final quark is completely negligible in this case
as well, and the difference between $\Gamma $ and $\Gamma _0$ for this
transition can be entirely attributed to the top mass.

\medskip 
{\begin{table}[tbp] \centering} 
\begin{tabular}{lll}
\hline\hline
\rule[2mm]{0mm}{2.5mm}Mode & $\Gamma _0\;\mathrm{(GeV)}$ & $\Gamma \;\mathrm{%
(GeV)}$ \\[0.5ex] \hline
\rule[2mm]{0mm}{2.5mm}$ s\rightarrow {d}\,\gamma \mathrm{\qquad }$ & $%
7.48\times 10^{-29}\qquad $ & $7.53\times 10^{-29}\qquad $ \\[0.5ex] 
$b\rightarrow s\,\gamma $ & $5.06\times 10^{-17}$ & $5.10\times 10^{-17}$ \\%
[0.5ex] 
$c\rightarrow {u}\,\gamma $ & $1.50\times 10^{-28}$ & $1.50\times 10^{-28}$
\\[0.5ex] 
$t\rightarrow c\,\gamma $ & $4.51\times 10^{-14}$ & $2.73\times 10^{-12}$ \\%
[0.5ex] \hline\hline
\end{tabular}
\caption{Widths for $q_i\to q_f \gamma$ transitions. 
\label{Width}} 
{\end{table}}

Before leaving this section we will make estimates of the branching fraction
for the $c\rightarrow {u}\gamma $ transition (i.e. its inclusive rate scaled
to that of the semileptonic decay $c\rightarrow ql^{+}\nu $). The correct
expression $\bar{F}=F_1-(1/3)F_2$ which we use differs in magnitude and sign
from the wrong combination $\bar{F}^{\prime }=-F_1-(1/3)F_2$ for the
relevant values of the argument $r_\ell $. The QCD-uncorrected branching we
have obtained, $B^{(0)}=3.90\times 10^{-16}$, is about 28 times larger than
an estimate \cite{burd} based on $\bar{F}^{\prime }$. To calculate the QCD
corrections, one begins with the Wilson coefficients evaluated at the $W$ mass
scale; in particular the coefficient 
\begin{equation}
c_7(M_{W})=-\frac 12\left[ \frac{\lambda _s}{\lambda _b}\bar{F}(r_s)+\bar{F}%
(r_b)\right]
\end{equation}
yields $2.065\times 10^{-3}$, to be compared with $-0.414\times 10^{-3}$
when it is calculated with $\bar{F}^{\prime }$. The Wilson coefficients are
then evolved down from $M_{W}$ to the renormalization scale $\mu =m_{c}$ to
give the effective coefficient $c_7^{\mathrm{eff}}(m_{c})$. 
The resulting QCD-corrected branching
fraction obtained is reduced by $1\%$ from Burdman et al.'s estimate \cite
{burd}. The reason for the smallness of this effect is that $c_7(M_{W})$
makes a very small contribution to $c_7^{\mathrm{eff}}(m_{c})$ which is
completely dominated by $c_2(M_{W})$. Nevertheless, the point being made is
that the correct input functions $F_p $ or, even better, $G_p$ and $H_p$ are
crucial for a proper evaluation of the Wilson coefficients.

\medskip 
{\begin{table}[tbp] \centering} 
\begin{tabular}{llll}
\hline\hline
\rule[2mm]{0mm}{2.5mm}Quark & $S=P\,(m_{c}=0)$ & $S\,(m_{c}\neq 0)$ & $%
P\,(m_{c}\neq 0)$ \\[0.5ex] \hline
\rule[2mm]{0mm}{2.5mm}{$d$\qquad } & $6.12\times 10^{-8}\quad $ & $6.20\times
10^{-8}\quad $ & $6.05\times 10^{-8}\quad $ \\[0.5ex] 
$s$ & $8.12\times 10^{-6}$ & $9.02\times 10^{-6}$ & $8.81\times 10^{-6}$ \\%
[0.5ex] 
$b$ & $5.95\times 10^{-3}$ & $6.00\times 10^{-3}$ & $5.90\times 10^{-3}$ \\%
[0.5ex] \hline
\rule[2mm]{0mm}{2.5mm}$\Gamma (\mathrm{GeV})$ & $2.7269\times 10^{-12}$ & 
\multicolumn{2}{l}{\qquad \qquad $\quad 2.7273\times 10^{-12}$} \\%
[0.5ex] \hline\hline
\end{tabular}
\caption{Contributions to $t\,\to c+\gamma$ in the linear
approximation in $m_c^2$ and $m_t^2$.\label{TabMod} } 
{\end{table}}

\section{MASS EFFECT IN\\ THE TOP QUARK DECAY}

Since $m_{c}\ll m_{t}$, and $G_p$ and $H_p$ must be comparable in magnitude,
the amplitude (\ref{RLAmp}) reduces to $RMG_p$ and we need only to perform
an accurate evaluation of the integrals 
\begin{equation}
G_p(r)=\int_0^1\!\mathrm{d}x\int_0^{1-x}\!\mathrm{d}y\,\frac 1{\Lambda_p}
\,g_p(r,x,y)\,,
\end{equation}
where $\Lambda _1=\,1-x+r\,x-r_i\,xy$ and $\Lambda _2=\,x+r\,(1-x)-r_i\,xy$,
with $r_i=r_{t}$, $r=r_\ell $, and $\ell =d, s, b$. After integration over $y
$, which yields logarithm functions, the $x$ integration leads to 
\begin{eqnarray}
G_1(r) &=&\frac{(2+r)}{2r_i}+\frac 1{r_i^2}[(r_i-1)(2-r)+r^2]\,J  \nonumber
\\
&&+\frac 1{r_i^2}[2(1-r_i)+r]\left[ \mathcal{L}_2(1-r)-\mathcal{L}_2\left(
\frac 1\alpha \right) -\mathcal{L}_2\left( \frac 1\beta \right) \right] \,,
\label{EqG1} \\
G_2(r) &=&\frac{-(2+r)}{2r_i }+\frac 1{r_i^2}[(r_i-1)(2-r)+r^2]\,J  \nonumber
\\
&&+\frac 1{r_i^2}[r(r_i-r-2)]\left[ \mathcal{L}_2\left( 1-\frac 1r\right) -%
\mathcal{L}_2\left( \frac 1{1-\alpha }\right) -\mathcal{L}_2\left( \frac
1{1-\beta }\right) \right] \,.  \nonumber \\
&&  \label{EqG2}
\end{eqnarray}
The parameters $\alpha ,\,\beta $ are defined by 
\begin{eqnarray}
\alpha &=&\frac 1{2r_i}\,(1+r_i-r-\sqrt{\Delta }\,),\qquad \beta =\frac
1{2r_i}\,(1+r_i-r+\sqrt{\Delta }\,),  \nonumber \\
\Delta &=&(1+r_i^2+r^2)-2(r_i+r+r_i\,r)\,.
\end{eqnarray}
For the transition under discussion, they satisfy the conditions $0<\alpha <
\beta <1$. $J$ stands for 
\begin{eqnarray}
J &=&-1+\ln r_i+\frac r{1-r}\ln r+\alpha \ln \alpha +\left( 1-\alpha \right)
\ln (1-\alpha )  \nonumber \\
&&+\beta \ln \beta +(1-\beta )\ln (1-\beta )\,+i\pi \frac{\sqrt{\Delta }}{r_i%
}\,,
\end{eqnarray}
and $\mathcal{L}_2$ is the dilogarithm (Spence integral)\cite{abra}
represented by 
\begin{equation}
\mathcal{L}_2(x)=-\int_0^x\frac{\ln (1-t)}t\,\mathrm{d}t\,,  \label{Spence}
\end{equation}
which admits the expansion series $\mathcal{L}_2(x)=\sum_{k=1}^\infty
x^k/k^2 $ for $\left| x\right| \leq 1$.

With a top mass larger than the mass of the $W$ boson and the internal quark, $%
\Lambda _1$ and $\Lambda _2$ may change signs over the range of the $x,\,y$
integrations, and hence the arguments of the various logarithm functions
that appear in the course of the integrations may become negative, and
imaginary parts will arise. Physically they signal the presence of the
intermediate states on the mass shell. Thus, for each internal quark $\ell
=d, s, b$, the imaginary parts of $G_p(r_\ell )$ correspond to the real
emission process $t\rightarrow W^{+}+q_\ell +\gamma $.

\medskip 
{\begin{table}[tbp] \centering} 
\begin{tabular}{lccc}
\hline\hline
\rule[2mm]{0mm}{2.5mm} & $G_1$ & $G_2$ & $S$ \\[0.3ex] \hline
\rule[2mm]{0mm}{2.5mm}$d$ & $0.84375-i\,0.80176$ & $%
-0.19858+i\,0.82598$ & $0.45497-{i\,}0.53855$ \\[0.5ex] 
$s$ & $0.84375-i\,0.80177$ & $-0.19860+i\,0.82600$ & $0.45498-{i\,}0.53855$
\\[0.5ex] 
$b$ & $0.84541-i\,0.80377$ & $-0.20823+i\,0.83341$ & $0.45742-{i\,}%
0.54079$ \\[0.5ex] \hline
\rule[2mm]{0mm}{2.5mm} & \multicolumn{3}{l}{\qquad \qquad \qquad \qquad
\qquad $\Gamma (\mathrm{GeV})=8.45\times 10^{-13}$} \\[0.5ex] \hline\hline
\end{tabular}
\caption{Values of $G_1$ and $G_2$ for $d,\,s$, and $b$ in
$t\,\to\,c\,+\gamma$.\label{G1G2}} 
{\end{table}}

Values of $G_1(r_\ell )$, $G_2(r_\ell )$, and $S(r_\ell )=\frac
12[G_1(r_\ell )-\frac 13G_2(r_\ell )]$ for relevant $r_\ell $ in $%
t\rightarrow c\gamma $ are recorded in Table \ref{G1G2}. They depend weakly
on the mass of the internal quark, being dominated by the large mass of the
initial top quark. Thus, the Glashow--Iliopoulos--Maiani mechanism works
effectively to suppress the amplitude 
\begin{equation}
A=B=\frac 12\sum_{\ell =d,s,b}\lambda _\ell \left[ G_1(r_\ell )-\frac
13G_2(r_\ell )\right] \,.  \label{Abg}
\end{equation}
In contrast, in the zeroth or first order approximation (in $r_i,\,r_f$) of
the transition amplitudes, the functions $S(r_\ell )$ and $P(r_\ell )$ 
depend strongly on the internal quark mass, as shown in Table \ref{TabMod}.
The contribution of the intermediate $b$ quark is then overwhelmingly
dominant and the GIM mechanism hardly operative. However, this conclusion
turns out to be erroneous, as is clear from the data presented in Table \ref
{G1G2}. Finally, the exact $G_1$ and $G_2$ scale as $r_{{t}}^{-1}\sim 1$
[cf. (\ref{EqG1})--(\ref{EqG2})], whereas their approximate values scale at
best as $r_{{t}}r_{{b}}\sim r_{{b}}\sim 10^{-2}$ [cf. (\ref{Ghp})--(\ref
{lastKL})], which explains the striking differences in magnitudes that we
see in Tables \ref{TabMod} and \ref{G1G2}.

With the amplitude given in (\ref{Abg}), we find the width $\Gamma ({t}%
\rightarrow c\gamma )\approx 8.45\times 10^{-13}$ GeV, which falls between
the values presented in Table 2 based on the approximation $r_i=
r_f= 0$ and the linear approximation (\ref{DecRat}). In brief, the
effect of the initial fermionic mass in $t\to c\gamma $ and, by extension, $%
\tau \to \mu \gamma $ is both subtle and substantial.

\section{CONCLUSIONS}

We have supplemented the Inami--Lim formula for the $s\to d\gamma $
effective vertex with an equally simple formula for the $c\to u\gamma $
vertex [Eq.~(\ref{right})] under the same assumptions that the external
fermionic masses are negligible. We then proceeded on to re-examine the
external mass effects and obtain the corresponding formulas [Eqs.~(\ref
{RLAmp}), (\ref{Ghp})--(\ref{lastKL})] valid to linear order in $M^2$ and $%
m^2$, the results of which are shown in Table 2. Since this approximation
breaks down (cf.~Table \ref{TabMod}) in the cases such as $t\to c\gamma $,
in which the initial fermion mass is much larger than any internal mass, we
derive the exact one-loop formula for such transitions [Eqs.~(\ref{EqG1})--(%
\ref{Spence})] and evaluate the corresponding rate of the $t\to c\gamma $
transition in Table 4.

Given the importance of flavor-changing electromagnetic transitions in
testing the standard model and probing new physics, it is essential to have
reliable results for these processes at the lowest, one-loop level since
they are the key components in building up more sophisticated calculations.

\bigskip \noindent{\bf ACKNOWLEDGMENTS}\par
The work of QHK was supported by the Natural Sciences and Engineering 
Research Council of Canada.

\newpage
\noindent FIGURE CAPTIONS

\begin{description}
\item  Fig. 1 Contributions to $q_i\rightarrow q_f\gamma $ vertex: emission
from bosons.

\item  Fig. 2 Contributions to $q_i\rightarrow q_f\gamma $ vertex: emission
from internal quark.

\item  Fig. 3 Emission from initial or final quark.

\item  Fig. 4 Kinematic variables in one-loop diagrams.
\end{description}

\end{document}